# Observation of Anisotropic Event Shapes and Transverse Flow in Au+Au Collisions at AGS Energy

J. Barrette[4], R. Bellwied[8], S. Bennett[8],

P. Braun-Munzinger[6], W. E. Cleland[5], M. Clemen[5],

J. Cole[3], T. M. Cormier[8], G. David[1], J. Dee[6],

O. Dietzsch[7], M. Drigert[3], S. Gilbert[4], J. R. Hall[8],

T. K. Hemmick[6], N. Herrmann[2], B. Hong[6], C. L. Jiang[6],

Y. Kwon[6], R. Lacasse[4], A. Lukaszew[8], Q. Li[8],

T. W. Ludlam[1], S. McCorkle[1], S. K. Mark[4], R. Matheus[8],

E. O'Brien[1], S. Panitkin[6], T. Piazza[6], C. Pruneau[8],

M. N. Rao[6], M. Rosati[4], N. C. daSilva[7], S. Sedykh[6],

U. Sonnadara[5], J. Stachel[6], H. Takai[1], E. M. Takagui[7],

S. Voloshin[5], G. Wang[4], J. P. Wessels[6], C. L. Woody[1],

N. Xu[6], Y. Zhang[6], Z. Zhang[5], C. Zou[6]

(E877 Collaboration)

[1] *Brookhaven National Laboratory, Upton, NY 11973*
[2] *Gesellschaft für Schwerionenforschung, Darmstadt, Germany*
[3] *Idaho National Engineering Laboratory, Idaho Falls, ID 83402*
[4] *McGill University, Montreal, Canada*
[5] *University of Pittsburgh, Pittsburgh, PA 15260*
[6] *SUNY, Stony Brook, NY 11794*
[7] *University of São Paulo, Brazil*
[8] *Wayne State University, Detroit, MI 48202*

Event shapes for Au + Au collisions at 11.4 GeV/c per nucleon were studied over nearly the full solid angle with the E877 apparatus. The analysis was performed by Fourier expansion of azimuthal distributions of the transverse energy ($E_T$) measured in different pseudorapidity intervals. For semicentral collisions a pronounced event anisotropy is identified beyond that expected due to fluctuations in particle multiplicity. The signal decreases for peripheral and very central collisions. The amplitude of the flow signal reaches up to 7% of the mean $E_T$.

Phenomenological analysis [1] of early data and later theoretical predictions [2] have led to a picture of nearly complete stopping of baryons in ultra-relativistic nuclear collisions at AGS energies. Recently, these predictions have been experimentally confirmed [3–6]. This stopping scenario implies energy and baryon densities of up to 10 times that of nuclei in their ground state [2,7]. Such large compression factors should yield pressure-induced flow effects [8,9], leading to a deflection of nucleons away from the high density region. The degree of these flow effects depends on the build-up of pressure gradients, on mean field interactions [9] and on the equation of state of the hot matter [8, 10]. Their study should provide insight into the properties of the dense medium formed in the collision. As a first step towards identifying such collective effects we report here the presence of pronounced event anisotropies indicating directed sidewards flow of the hot and excited matter formed in Au+Au collisions at AGS energy. The results were obtained by analyzing, in different pseudo-rapidity ($\eta$) intervals, distributions in azimuthal angle ($\phi$) of transverse energy ($E_T$), measured with nearly $4\pi$ coverage in calorimeters surrounding the target.

Following predictions based on hydrodynamical considerations [11] flow effects in nucleus-nucleus collisions were first observed and systematically studied at beam energies of a few hundred MeV/nucleon [12]. The present investigations are focussed on much higher energy where a phase transition from hadronic to quark-gluon matter is possible. The sensitivity of flow [8, 10] to the equation of state implies that our measurements may provide information complementary to the study of single particle observables and two-particle correlations generally used in the search for quark matter [13].

First results for the production of transverse energy in collisions of 11.4 A GeV/c $^{197}$Au beams with nuclear targets at the BNL AGS accelerator have been reported previously [4]. This Au + Au data set, obtained with two calorimeters, the target

calorimeter(TCal, -0.5 < $\eta$ < 0.8) and the participant calorimeter (PCal, 0.83 < $\eta$ < 4.5), was used for the flow analysis. Both detectors cover $2\pi$ azimuthally, and consist of 832 NaI crystals for TCal, and 512 towers for PCal with individual readout. Events were collected with four levels of PCal $E_T$ triggers. Details of the experimental technique and the detector set-up can be found in [4]. The possibility to use the calorimetric technique without particle identification for flow measurements has been investigated [14] for the system Si+Pb. There it is shown that calorimetric measurements are a sensitive tool to study flow phenomena.

For the event shape analysis, where the absolute calibration of the transverse energy scale is not important, we have not used leakage and response correction as done previously [4, 15]. Rather, emphasis was placed on very careful gain equalization requiring azimuthal symmetry of the energy deposit averaged over many events. In addition, to correct for "dead" channels and other instrumental asymmetries, we required that the centroid of the first and second harmonic (for definition see eq. (1) below) of the azimuthal $E_T$ distribution be zero on average. Our event shape analysis was motivated by a concept for flow analysis tailored to ultrarelativistic collisions [16]. A more general approach based on Fourier expansion of the azimuthal distribution[17] is employed in our data analysis.

The E802 collaboration has previously reported [18] a correlation between particle production ratios and transverse momenta of projectile fragments (spectators) for Si + Au collisions. The pseudorapidity coverage of our detectors is such that target and projectile spectator nucleons account for negligible contributions to the energy deposit. Thus the event asymmetries reported here are carried by particles from the hot and dense matter in the collision region.

To study the centrality dependence of a possible flow signal, events were sorted into 12 bins, each 20 GeV wide, from 40 to 280 GeV according to the transverse energy deposited in PCal (PCal $E_T$). Corrections for interactions occurring upstream of the target were performed using data taken with an empty target frame. The correction is important only at low $E_T$, amounting to 39 % in the 80 - 100 GeV interval, 14% in the 120 - 140 GeV interval, and negligible for larger $E_T$ values. After background cuts there are about 80,000 events left for the analysis. A possible dependence of the event shapes on the angle of the incoming beam was studied using the beam vertex detector and found to be negligible.

For the analysis the pseudorapidity space was subdivided into three intervals: backward ( -0.5 < $\eta^b$ < 0.8 ), middle (0.83 < $\eta^m$ < 1.85), and forward (1.85 < $\eta^f$ < 4.7). Central (pseudo-)rapidity is about 1.7 [4]. The somewhat asymmetric choice of intervals is dictated by detector boundaries and statistics considerations. To study the event shapes the measured azimuthal distributions were decomposed into normalized Fourier components [17]:

$$v_n e^{in\psi_n} = \frac{\int_0^{2\pi} \varepsilon_T(\phi) e^{in\phi} d\phi}{\int_0^{2\pi} \varepsilon_T(\phi) d\phi} = \frac{\sum_j \varepsilon_T^j e^{in\phi_j}}{\sum_j \varepsilon_T^j}, \qquad (1)$$

where $\phi_j$ is the azimuthal angle of the $j^{th}$ detector channel and $\varepsilon_T^j$ is the transverse energy measured in that channel. Fourier components $n = 1, 2, 3, 4$ were studied. For $n = 1$, $v_1$ describes the radial displacement of the distribution, while $\psi_1$ is the angle, relative to the (horizontal) x-axis, of the reaction plane spanned by the impact parameter vector and the beam direction. Nonzero values for higher harmonics would indicate more complex events shapes ($n = 2$ indicates elliptic components $etc.$). The main point of the following analysis is to determine which of the Fourier components $v_n$ are significantly different from zero and to study their centrality or $E_T$ dependence. Because of the reduced fluctuations in transverse energy production for the Au+Au system [4], the transverse energy is closely anti-correlated with the impact parameter.

In Figures 1(a,b) we display distributions of the orientation of the reaction plane, $\psi_1$, for the backward and forward pseudorapidity intervals for events with PCal $E_T$ in the range $E_T = 200 - 220$ GeV. Both spectra are flat as expected, indicating the quality of the symmetrization and gain equalization described above. The sensitivity of our data to the reaction plane is evident from Fig. 1(c) where the distribution of the angular difference $\Delta\psi_1^{bf} = \psi_1^b - \psi_1^f$ shows a pronounced forward-backward correlation with a peak at $\Delta\psi_1^{bf} = \pi$. To quantify the degree of correlation we define the ratio

$$R = \frac{N(|\Delta\psi_1^{bf} - \pi| < \pi/2)}{N(|\Delta\psi_1^{bf} - \pi| > \pi/2)}. \tag{2}$$

This ratio $R$ depends strongly on transverse energy, as shown in Figure 2. It reaches its maximum for intermediate centrality, but approaches 1 (implying isotropy) for peripheral and the most central collisions. This implies that the best determination of the reaction plane is for events of intermediate centrality. Due to larger leakage fluctuations for the TCal the best resolution in the reaction plane measurement comes from data using only the PCal. Estimates following a procedure described in [17] with the shape parameters determined below imply a resolution of $\sigma_{\psi_1} \approx 30°$.

In the absence of flow, the central limit theorem ensures a Gaussian distribution for $d^2N/dv_n^2$ centered at zero, i.e.

$$\frac{1}{N}\frac{dN}{v_n dv_n} = \frac{\exp\left(-\frac{v_n^2}{2\sigma^2}\right)}{\sigma^2}, \quad \sigma = \frac{1}{\sqrt{M}}\sqrt{\frac{\langle\epsilon_T^2\rangle}{2\langle\epsilon_T\rangle^2}}. \tag{3}$$

Here, $N$ is the number of events, $\epsilon_T$ is the transverse energy per particle, and $M$ is the mean multiplicity of particles in the $\eta$ interval for events in a given centrality bin. The variance of the Gaussian, $\sigma$, is independent of $n$ but is determined by $M$ and the $\epsilon_T$ distribution. The presence of flow will result in a shift by $\tilde{v}_n$ of the centroid of the $d^2N/dv_n^2 d\psi_n$ distribution. Assuming that flow does not change the variance $\sigma$ one can integrate over $\psi_n$ to obtain

$$\frac{1}{N}\frac{dN}{v_n dv_n} = \frac{1}{\sigma^2} \exp\left(-\frac{\tilde{v}_n^2 + v_n^2}{2\sigma^2}\right) I_0\left(\frac{\tilde{v}_n v_n}{\sigma^2}\right), \qquad (4)$$

where $I_0$ is the modified Bessel function. The distribution specified by eq. (4) will exhibit a minimum at $v_n = 0$ if $\tilde{v}_n > \sqrt{2}\sigma$. A sufficient condition for anisotropic event shapes is, consequently, the observation of such a minimum in the data. However, a minimum at zero is not required for the extraction of $\tilde{v}_n$.

For each of the three pseudorapidity intervals the measured transverse energy data were therefore converted into $dN/v_n dv_n$ distributions for $n = 1, 2, 3$ and $4$. The results are shown for the forward pseudorapidity interval in Figure 3 for five centrality bins, ranging from peripheral to very central collisions. The most dramatic sequence of shapes is observed for $n = 1$, where a pronounced minimum at $v_1^f = 0$ is observed in the intermediate centrality intervals, with maxima at zero for the most central and the most peripheral intervals. This demonstrates unambiguously the existence of directed, sidewards flow in semi-central Au + Au collisions at the AGS.

The curves in Fig. 3 are the results of fitting eq. (4) to the data, with 4 free parameters to fit, simultaneously, the 4 distributions obtained at each centrality. The parameters are $\tilde{v}_n$ ($n = 1 - 4$) and $\sigma$. The parameter $\tilde{v}_3$ should vanish for symmetric systems. Our results are consistent with this expectation. To reduce the number of parameters in the fit we enforced $\tilde{v}_3$ to be zero. The resulting anisotropy parameters $\tilde{v}_n^b$, $\tilde{v}_n^m$ and $\tilde{v}_n^f$ for the three $\eta$ intervals are plotted as function of centrality in Figure 4 for $n = 1, 2, 4$. From Figs. 4(a) and (c) one can see that both $\tilde{v}_1^b$ and $\tilde{v}_1^f$ are nonzero over a significant centrality range, and both reach their maximum values at intermediate centrality. The shifts go into opposite direction in the backward and forward hemisphere, as demonstrated already in Figs. 1 and 2. These results demonstrate that, in the backward and forward region, the azimuthal $E_T$ distribution in each event is shifted off center. Since we use normalized Fourier coefficients the magnitude of $\tilde{v}_1$ shown in Fig. 4 implies that the amplitude of the directed transverse energy flow reaches 7 %. The magnitude of the directed flow gradually decreases towards very central or very peripheral collisions. The flow signals do not fully vanish even for the most central collisions, as expected because of fluctuations between the impact parameter and any global observable.

We have used the fit parameters $\tilde{v}_1^b$, $\tilde{v}_1^f$ and $\sigma$ to calculate the ratio $R$ in eq. (2), assuming that there is no other correlation except flow. The results of this calculation are plotted as the histogram in Fig. 2 and demonstrate the internal consistency of the data.

In Fig. 4(b), the fit parameters obtained for the $\eta^m$ interval, *i.e.* near central rapidity, are plotted. As expected in this rapidity interval the $n = 1$ term corresponding to directed flow is always close to zero. Small but nonzero $\tilde{v}_2$ terms appear in all rapidity intervals, indicating that the distribution has a slightly elliptical shape [16, 17]. Analysis similar to that shown in Fig. 1 shows that the long axis lies in the reaction plane. The origin of this elliptic shape is not yet known. It may lie in the variation with pseudorapidity of the shift $\tilde{v}_1$ within our finite bin size. Since our mid-pseudorapidity bin is not symmetric about central pseudorapidity, we would expect, in this case, a nonzero $\tilde{v}_1$, not seen in the data. Elliptic event shapes near mid-rapidity may also originate from the pressure gradient during the collision (squeeze-out). In that case the long symmetry axis should point perpendicular to the reaction plane, not in agreement with the data. Assuming that the elliptic shape is due to the rapidity variation of the $n = 1$ term a small perpendicular squeeze-out should be visible as a nonzero $\tilde{v}_4$. This is actually observed in this pseudorapidity bin although it is very small.

In conclusion, we have demonstrated that using calorimetric methods one can determine the orientation of the reaction plane for Au + Au collisions at AGS energy with rather good precision. For the first time we have observed directed flow for ultrarelativistic Au + Au collisions. Higher harmonics in the event shape were also detected but their origin is unclear, at present.

We thank the AGS staff, W. McGahern and Dr. H. Brown for excellent support and acknowledge the untiring efforts of R. Hutter in all technical matters. Financial support from the US DoE, the NSF, the Canadian NSERC, and CNPq Brazil are gratefully acknowledged. One of us (JPW) thanks the A. v. Humboldt Foundation for support.

# REFERENCES

[1] J. Stachel and P. Braun-Munzinger, Phys. Lett. **B216**, 1(1989).

[2] H. Sorge, A. von Keitz. R. Mattiello, H. Stöcker, W. Greiner, Phys. Lett. **B243**, 7(1990).

[3] J. Barrette *et al.*, E814 collaboration, Phys. Rev. Lett. **64**, 1219(1990).

[4] J. Barrette *et al.*, E814 Collaboration, Phys. Rev. Lett. **70**, 2996(1993).

[5] F. Videbaek, E802 Collaboration, Proc. Int. Workshop "Heavy Ion Physics at the AGS '93", G. Stephans, S. Steadman, W. Kehoe, eds., MITLNS-2158, p. 89.

[6] S. Eiseman *et al.*, E810 Collaboration, Phys. Lett. **B292**, 10(1992).

[7] Y. Pang, T. J. Schlagel, and S. K. Kahana, Phys. Rev. Lett. **68**, 2743(1992); S. H. Kahana, Proc. Int. Workshop "Heavy Ion Physics at the AGS '93 " (Ref. [5] , p. 263)

[8] M. Hofmann et al., Nucl. Phys. **A566**, 15c(1994).

[9] R. Mattiello, A. Jahns, H. Sorge, H. Stöcker, W. Greiner, UFTP preprint 352/1994, Phys. Rev. Lett. (submitted).

[10] L. V. Bravina, N. S. Amelin, L. P. Csernai, P. Levai and D. Strottmann, Nucl. Phys. **A566**, 461c(1994).

[11] W. Scheid, H. Müller, W. Greiner, Phys. Rev. Lett. **32**, 741(1974).

[12] H. H. Gutbrod, A. M. Poskanzer, H. G. Ritter, Rep. Prog. Phys. **52**, 1267(1989).

[13] For a recent review of both CERN and AGS data see J. Stachel and G. R. Young, Annu. Rev. Nucl. Part. Sci. **42**, 237(1992) and references quoted there.

[14] A. Gavron, Nucl. Instr. Meth. **A273**, 371(1988).

[15] Z. Zhang, P. Braun-Munzinger, W. E. Cleland, G. David, D. Lissauer, in print Nucl. Instr. Meth. (1994).

[16] J. Ollitrault, Phys. Rev. **D46**, 229(1992) and Phys. Rev. **D48**, 1132(1993).

[17] S. Voloshin and Y. Zhang, to be published.

[18] T. Abbott *et al.*, E802 collaboration, Phys. Rev. Lett. **70**, 1393(1993).

FIGURES

FIG. 1. Distribution of orientations of the reaction plane for events with (200–220) GeV $E_T$ in the PCal using the backward and forward pseudorapidity windows $\eta^b$ (a) and $\eta^f$ (b); (c) reaction plane angle difference distribution. The hatched regions indicate the boundaries used to calculate the ratio $R$ (see text).

FIG. 2. Centrality dependence of the backward-forward correlation $R$. The histogram was calculated using the anisotropy parameters $\tilde{v}_1^b$, $\tilde{v}_1^f$ and $\sigma$. The quantity $\sigma_{E_T} = \int_{E_T}^{\infty} \frac{d\sigma}{dE_T'} dE_T'$ (top scale) indicates the centrality. For details see text.

FIG. 3. Distributions of experimental parameters corresponding to harmonics $n = 1$–$4$ (points) and fits using eq (4) for the $\eta^f$ interval. Each row is labelled by the PCal $E_T$ bin. For details see text.

FIG. 4. Anisotropy parameters $\tilde{v}_1$ (○), $\tilde{v}_2$ (◇) and $\tilde{v}_4$ (□) in the three pseudorapidity intervals plotted as function of PCal $E_T$. The top scale is defined in Fig. 2.

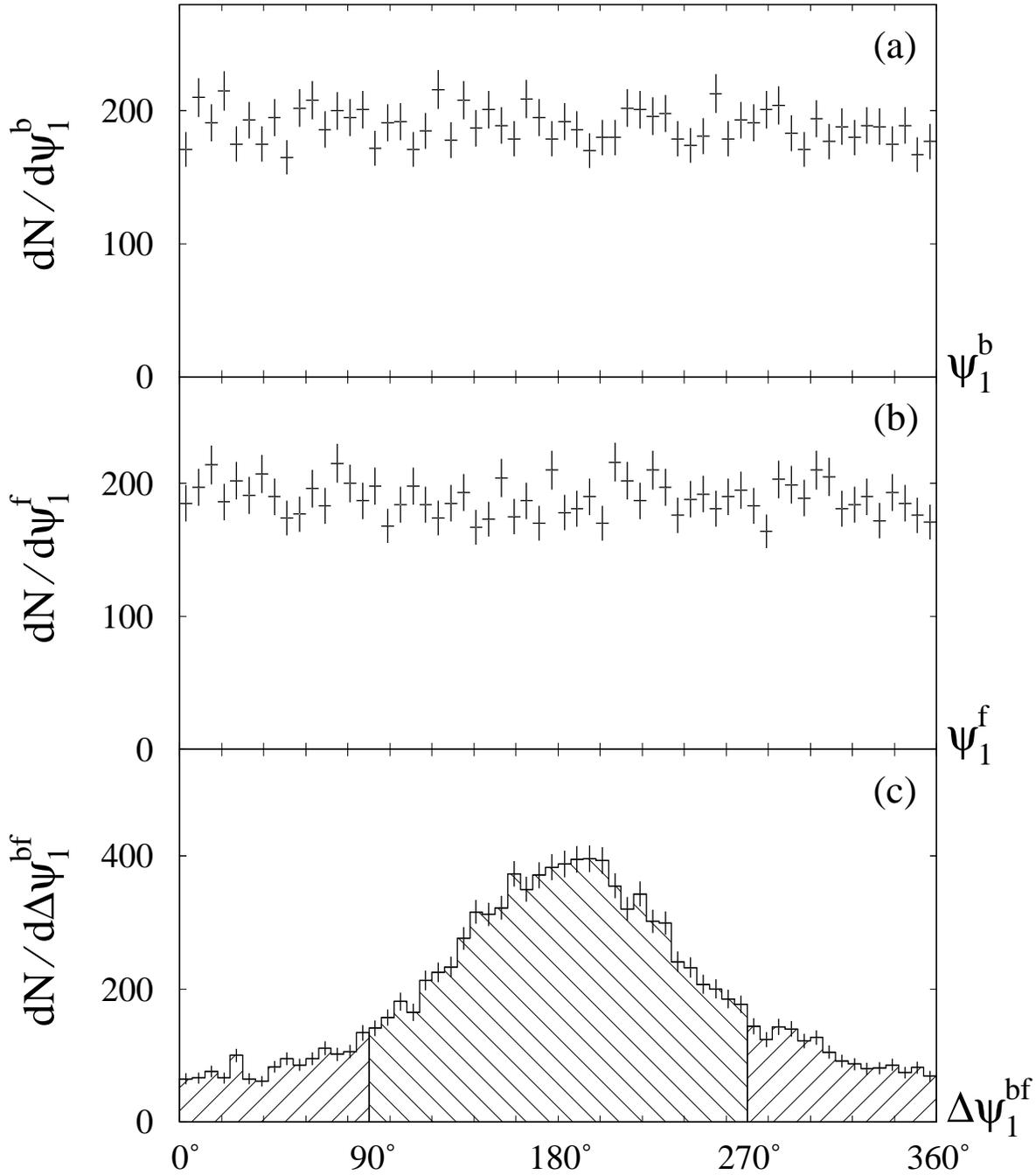

Fig. 1

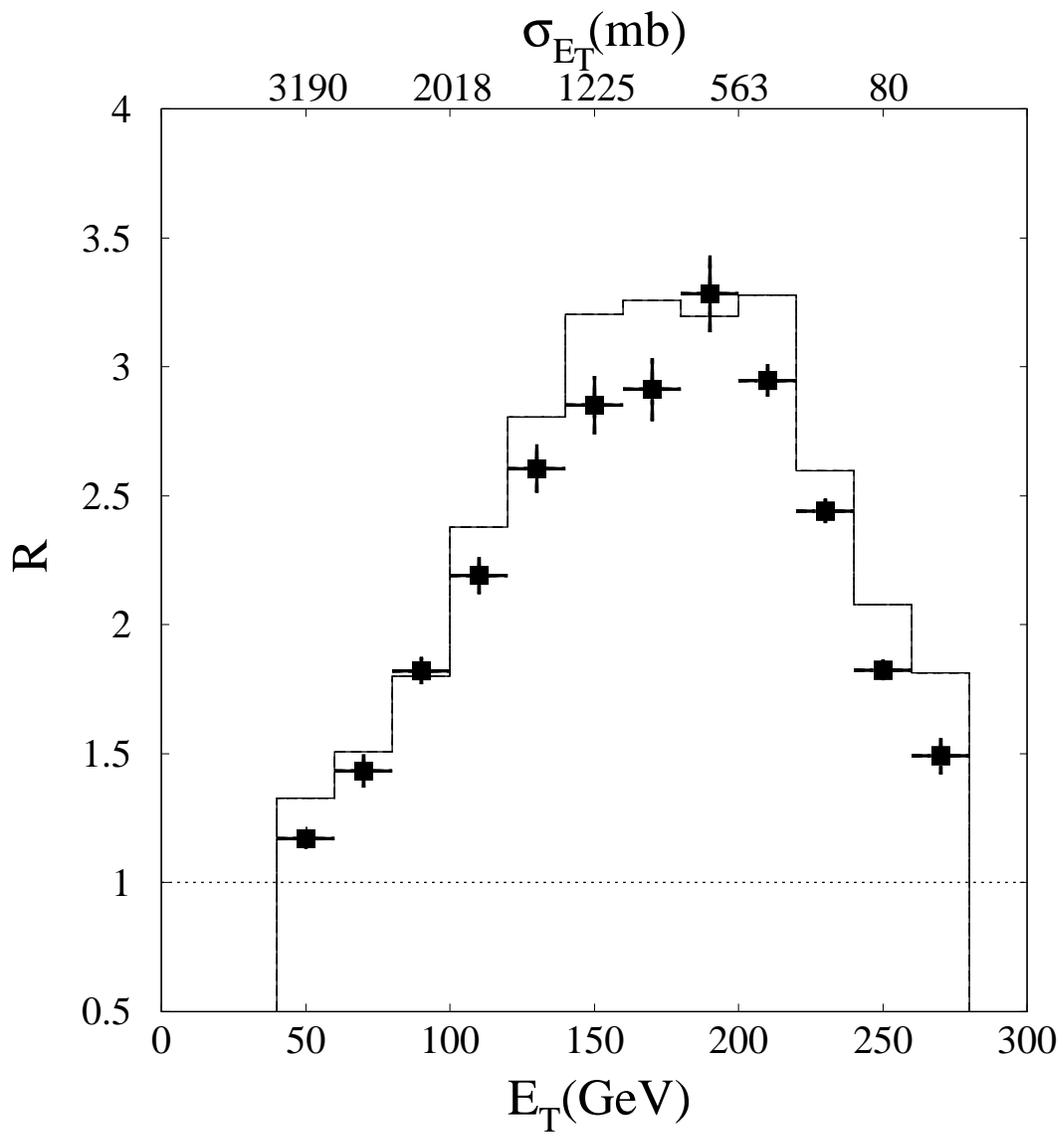

Fig. 2

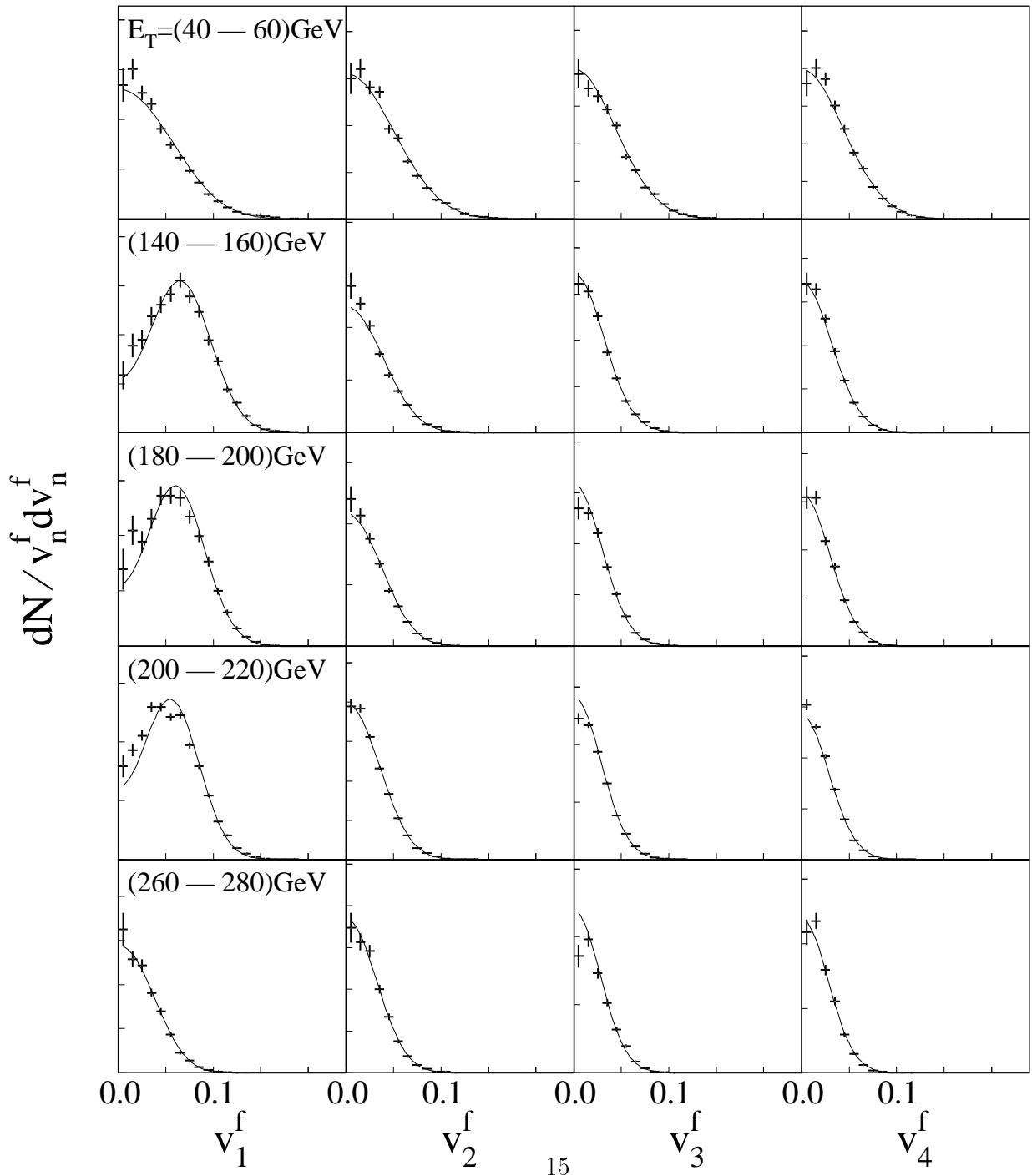

Fig. 3

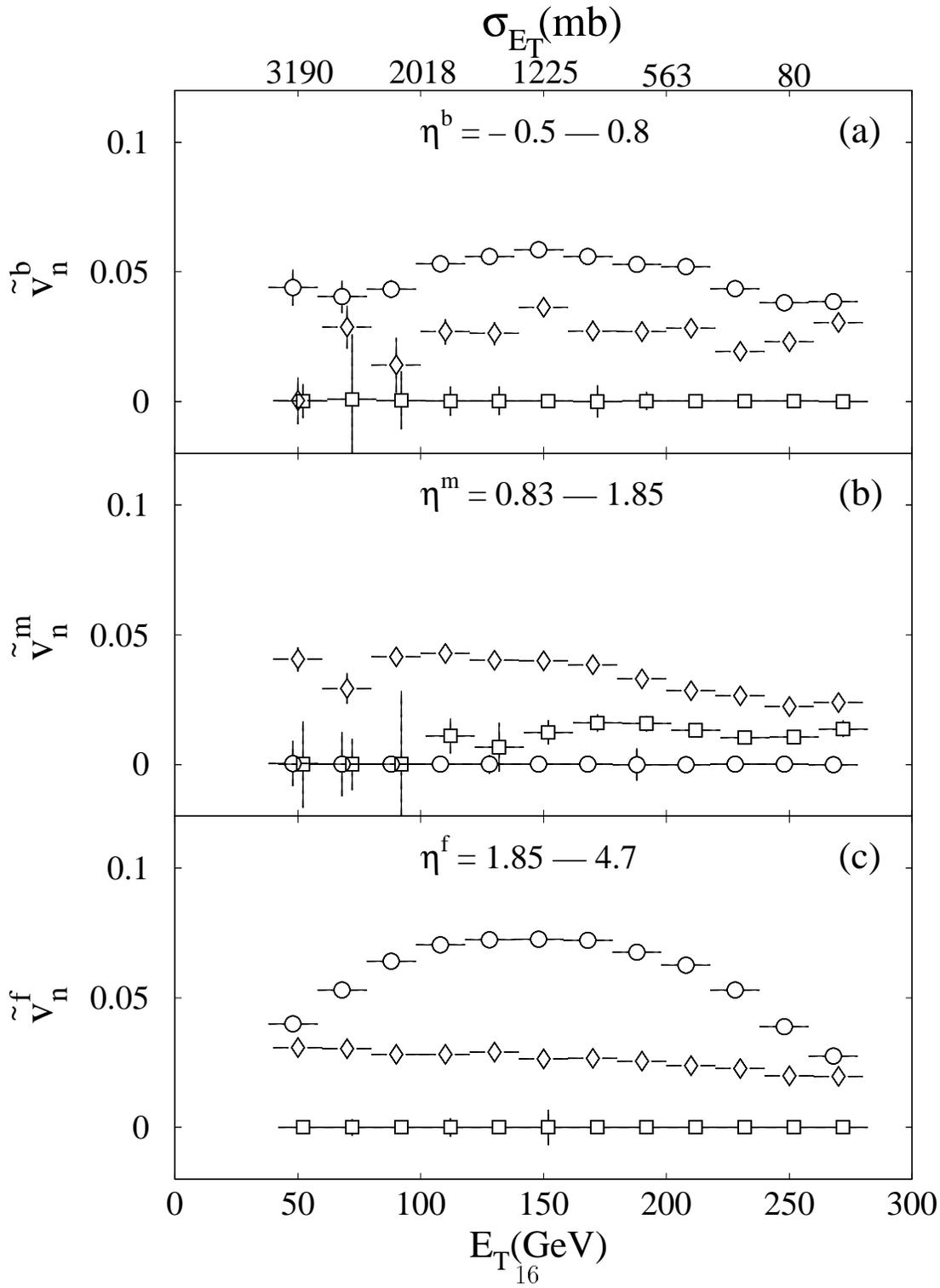

Fig. 4